 \documentclass[final,3p,times]{elsarticle}

 \usepackage{graphics}

\usepackage{amssymb}

\journal{Phys. Lett. B}

\begin{document}

\begin{frontmatter}







\title{Scaling Violation and Inelasticity of Very High Energy Proton-Proton Interactions.}%

\author{Tadeusz Wibig}


 \address{Physics Department,
University of {\L }\'{o}d\'{z}; \\
The Andrzej So\l tan Institute for Nuclear Studies,
Uniwersytecka 5, 90-950 \L \'{o}d\'{z}, Poland.}ź

\begin{abstract}
The pseudorapidity measurements at LHC, although in the central region only,
allows to perform preliminary tests of the multiparticle production 
extrapolation formula inspired by the recent cosmic ray data analysis. 
Feynman scaling violation in the form proposed originally 
by Wdowczyk and Wolfendale in 70s has been applied to the Pierre Auger Observatory 
and the Hi-Res group measurements. The consistency of the Extensive Air Shower development
and anisotropy data was found for smoothly rise of the scaling violation
parameter.
We have shown that the longitudinal momenta of produced particles determined inclusively as 
rapidity (pseudorapidity) distributions measured by LHC experiments 
follow the some universal high energy distribution 
scaled respectively. The high degree of Feynman scaling violation is confirmed. 
The decrease of the very high energy interaction inelasticity suggested by cosmic ray data
analysis
is found to be consistent with LHC measurements up to 7 TeV. 
\end{abstract}


\end{frontmatter}


\section{Introduction}
The inclusive description of minimum bias LHC events is not as spectacular as, e.g., 
Higgs hunting, but is essential for other very important scientific endeavours. 
One of them is the Ultra High-Energy Cosmic Ray (UHECR) problem and the answer to the
question of an existence of Greizen-Zatsepin-Kuzmin (GZK) cut-off \cite{gzk}. 
The origin and nature of cosmic rays is studied for almost exactly 100 years.
The great experimental effort has been taken recently by two groups: 
the Pierre Auger Observatory \cite{pao} and the Hi-Res experiment \cite {hires}.
The progress is observed, but the answers are still not decisive.
The cosmic rays
of energies of about $10^{20}$eV, if they are protons, should not reach us from 
cosmological distances. On the other hand anisotropy measurements show that they 
probably actually do. Our
knowledge about the nature of UHECR is based on observation of giant Extensive Air Showers 
(EAS) - cascades of secondary particles created in the atmosphere when the 
single atomic nucleus (proton in a simplest case) enters from above. It is expected that the
EAS initiated by protons and iron nuclei should differ. This difference is
determined by the rate of energy dissipation. Thus it depends strongly  on the
distribution of secondaries produced in the forward direction and on the nature of 
primary particle: its atomic mass.
The long-lasting discussions on the primary cosmic ray mass 
composition at the very end of the cosmic ray energy spectrum, in the 
so-called "ankle" region ( $E_{\rm lab} > 10^{18}$~eV), could 
not be conclusive also because of the lack on the more exact knowledge of 
the very high energy interaction physics, what makes the importance of the high
energy proton fragmentation even greater
for cosmic ray physicist, astronomers and cosmologists.

Searching for regularities and phenomenological description of the multiparticle production model
is as old as the modeling in high-energy physics itself. Starting from simple Fermi
thermodynamical model, to the first parton (quark) model propositions by 
Feynman, the model extrapolation
to much higher, cosmic ray energies was one of the most important and most wanted model predictions. 
It is usually in the form of a kind of scaling. 
The idea of limited fragmentation \cite{limi-fra} applied to the quark-jet 
hadronization led to introduction of the Feynman scaling variable of $x_F$ and the 
universal fragmentation function $f(x_F,\;s)=f_F(x_F)$
 \cite{feynman}. This brilliant idea works well for the
first collider experiments up to $\sqrt{s}\sim 60$~GeV. However, when applied to 
cosmic ray EAS development, it was questioned already at the "knee" energies of 
$E_{lab}\sim 10^{15}$~eV. The SPS ($\sqrt{s}\sim 200 - 900$~GeV) 
experiments allow to quantify the scaling violation.
The scale-breaking model of Wdowczyk and Wolfendale has been proposed to 
described the CR data at the beginning of '70 \cite{ww}. It is, in a sense, 
a generalization of the Feynman scaling idea introducing the one 
scaling violation parameter.

In Ref.~\cite{twphlww} 
we have shown that the light composition suggested by the studies 
of the anisotropy and the average depth of the shower maximum 
($x_{\rm max}$) does not contradict other results, mainly the width 
of the $x_{\rm max}$ distribution, only if one assume strong Feynman 
scaling violation. 

The rapidity (pseudorapidity) distributions were measured by LHC 
experiments: ALICE\cite{alice}, CMS\cite{cms900,cms7} and 
ATLAS \cite{atlas} (the last for $p_\bot >0.5$~GeV only) 
in the central rapidity region 
$| \eta | \lesssim 2.5$ for c.m.s. energies  
of 900 GeV, 2.3 TeV and 7 TeV.
Narrow range of a rapidity (pseudorapidity) at first sight does 
not allow to study important characteristics of very forward 
particle production. To study the fragmentation region 
new measurements, specially by much forward detectors (LHCf),
are welcome. But, as it will be shown below, 
the existing data can be used to test the 
scaling violation picture found in UHECR physics domain.

\section{Rapidity distribution}

Rapidity distributions measured in LHC experiments cover the central region where the
produced particles are dynamically separated from the valence quarks of
colliding hadrons. The central rapidity 
density $\rho(0) = 1 / \sigma  \left. \left( {\rm d \sigma / {\rm d} y} \right) \right|_{y=0}$ 
is the variable describing the particle production there. 
The original Feynman scaling preserves the value of the central rapidity density.
The plateau in rapidity is characteristic feature of independent jet fragmentation
model as well as statistical models with limited transverse momentum phase space.
Unfortunately, it is known for long, that such simple picture does not work. 

The phenomenological fit of the $\rho(0)$ rise made more than twenty years 
ago in Ref.~\cite{alner} is still valid. The 900 GeV LHC measurements match well 
SPS UA5 result. The systematic discrepancy seen by CMS detector  
\cite{cms900} does not change this general opinion. 

\subsection{Feynman scaling}

Feynman scaling \cite{feynman} can be expressed introducing one universal function
$f_F$ of the variable $x = p_\| / p_{\rm max}$ which describes the invariant momentum (longitudinal $p_\|$)
distribution of particles crated in the high-energy inelastic (and non single diffractive) interaction
\begin{equation}
{E \over {\sqrt{s}/2}}~ {1 \over \sigma }~{{d^3 \sigma} \over {d x \: d^2 p_\bot} }
~=~f(x,\:p_\bot,\: s)~=~f_F(x,\:p_\bot)
\label{xf}
\end{equation}
\noindent
where $\sqrt{s}$ is the interaction c.m.s. energy, $E$, $p_\|$ and 
$p_\bot$ are energy, and longitudinal and transverse momenta of 
outgoing particles ($p_{\rm max}\approx \sqrt{s}/2$). Change of variable from Feynman $x$ to rapidity $y$ gives
\begin{equation}
 {1 \over \sigma }~{{d^3 \sigma} \over {d y \: d^2 p_\bot} }~=~f_F \left( x(y),\: p_\bot \right)
\label{ysc}
\end{equation}
where $x(y)\:=\: \sqrt{p_\bot^2+m^2} / (\sqrt{s}/2) \sinh (y)$.
Using an approximate relation 
$\sqrt{p_\bot^2+m^2} \sinh (y) \approx p_\bot \sinh (\eta)$ and introducing the
very convenient variable: pseudorapidity $\eta= - \ln \tan ({\Theta/2})$ we have  
\begin{equation}
{1 \over \sigma }~{{d^3 \sigma} \over {d \eta \; d^2 p_\bot}}~=~f_F \left( {{2 p_\bot}\over {\sqrt{s} }}
\sinh (\eta)\;,p_\bot \right)~ ~.
\label{etasc}
\end{equation}
The integration over all $p_\bot$ is obvious with uncorrelated $p_\bot$ and $p_\|$ and the 
universality of the $p_\bot$ distribution
\begin{equation}
{1 \over \sigma }~{{d \sigma} \over {d \eta} }~=~
F_F \left( {{2 \langle p_\bot \rangle} \over {\sqrt{s} }} \sinh (\eta)\:\right) ~~.
\label{fsc}
\end{equation}
The factor  ${\langle p_\bot \rangle}$ is a constant related to the 
transverse momentum scale.

\begin{figure}
\centerline{
\includegraphics[width=7.2cm]{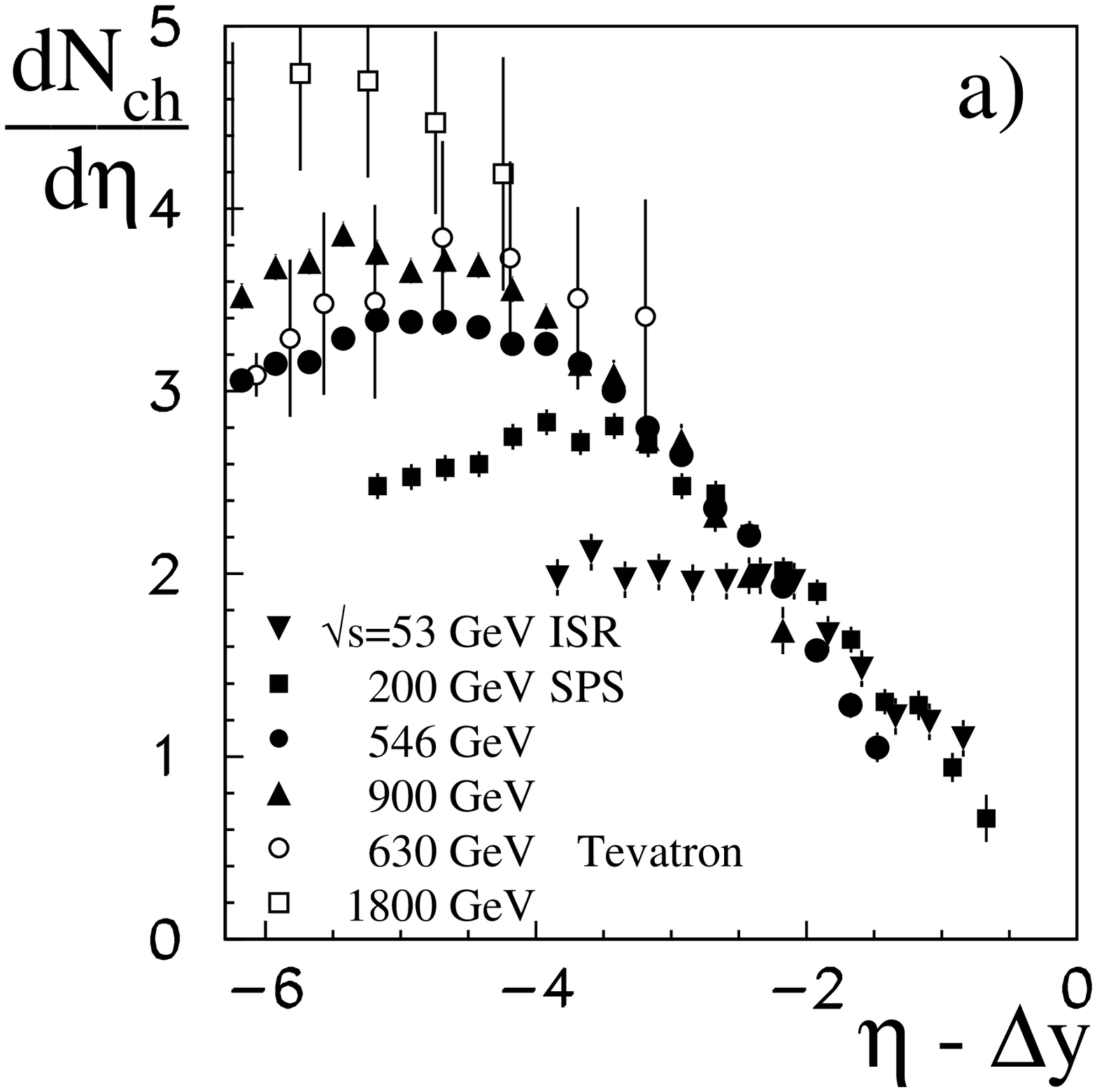}
\includegraphics[width=7.2cm]{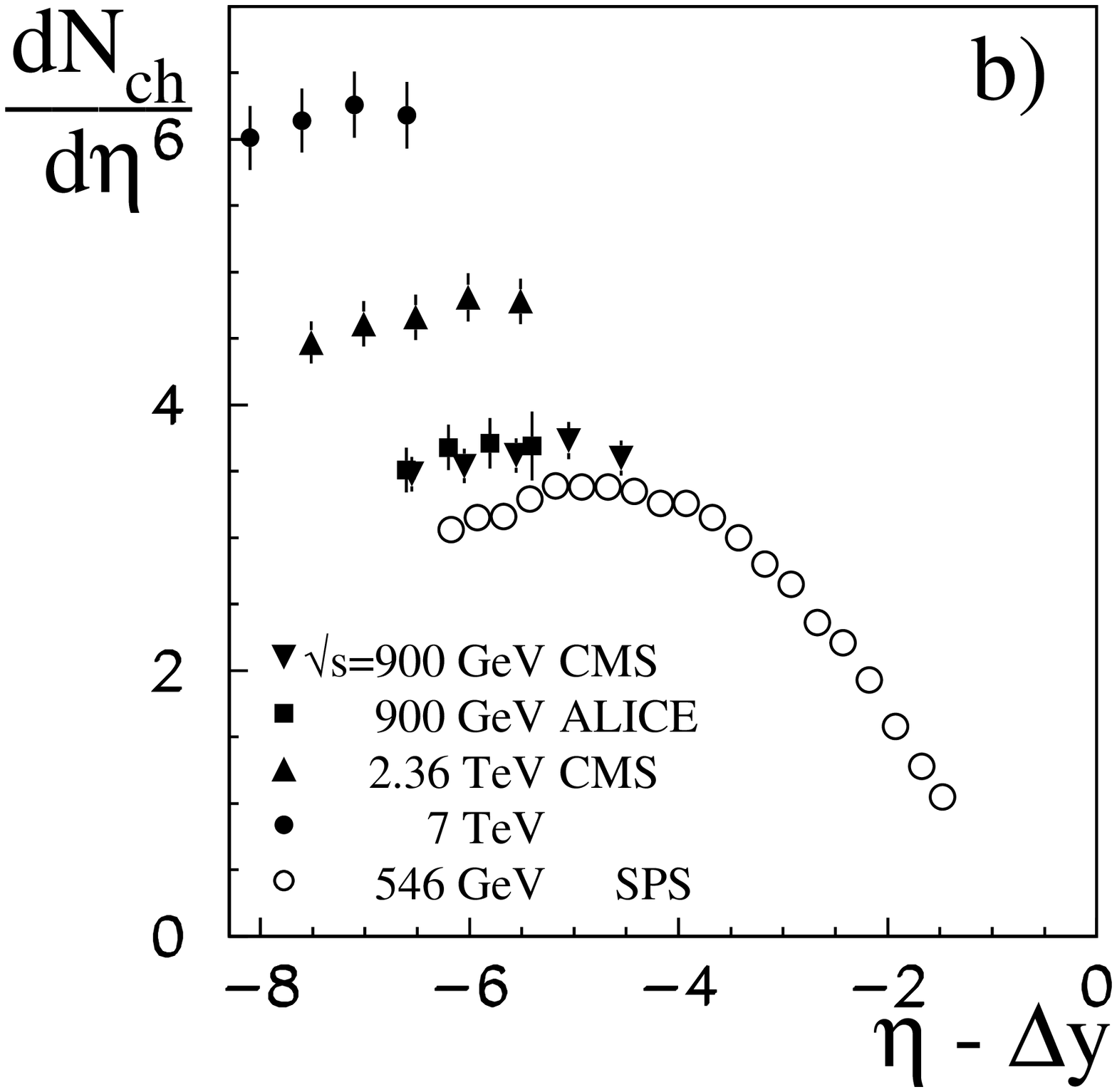}}
\caption{Pseudorapidity distributions 
shifted by $\Delta y=\ln (\sqrt{s} / m )$  for ISR, SPS and Tevatron measurements (a), and distributions
measured by LHC experiments at energies from 900 GeV to 7 TeV compared with 
SPS $\sqrt{s} = 546$ GeV UA5 result (b).}
\label{f1}
\end{figure}

We are interested of the extremely forward 
part of the (pseudo)rapidity distribution -- projectile fragmentation region. 
It is convenient to move the longitudinal momentum distribution to the anti-laboratory frame 
($\eta \rightarrow \eta '$)
where the projectile is at rest prior to the collision. This is done shifting the c.m.s. (pseudo)rapidity 
distribution by $\Delta y = \ln \:(\sqrt{s}/m)$
\begin{eqnarray}
~ \sinh ( \eta' ) 
~= ~
\sinh \left( \eta \:-\: \Delta y \right) ~= ~
\sinh \left( \eta \:-\: \ln (\sqrt{s}/m)\right) 
~\approx \nonumber\\
~{\rm e}^{\eta \:-\: \ln (\sqrt{s}/m)} 
/2 ~=~
 {{\rm e}^{\eta} \over 2} \: {m \over {\sqrt{s}}}  ~\approx~ 
 { m \over \sqrt s} \:
\sinh ( \eta )~~.
\end{eqnarray}
\noindent
After such transformation the direct comparison of particle production at different 
values of interaction c.m.s. energy is possible 
\begin{equation}
{1 \over \sigma }~{{d \sigma} \over {d \eta '}}~ \approx~ 
F_F \left({{2 \langle p_\bot \rangle} \over m }\:\sinh (\eta ')  \right)~=~
F_\eta \left(\eta '\right)~~.
\label{fscprim}
\end{equation}
This form of Feynman 
scaling was tested e.g. in ref.~\cite{alner} and it is found that it is valid only very approximately.
We can see this in Fig.~\ref{f1}a, where previous millennium data are plotted as a function of the 
anti-laboratory pseudorapidity. The recent data from CMS \cite{cms900,cms7} and ALICE \cite{alice} are shown in Fig.~\ref{f1}b.

It is known that Feynman scaling is 
violated at least by the continuous increase of 
the central rapidity density what is easily seen in 
Fig.~\ref{f1}. 

\subsection{Feynman scaling violation}

The original Feynman scaling implies that the inelasticity of proton-proton interaction, 
defined as a fraction of incoming energy carried by newly created particle,
is universal, the same 
for all interaction energies. The first observations suggested an attractive value of 0.5.
The rise of some characteristics of the interactions
(like, e.g., average $p_\bot$ or central rapidity density we mentioned above)  
makes the assumption about the constancy of the 
inelasticity not quite well justified. 
Introducing the multiplicative factor proportional to the observed rise of the 
rapidity plateau to the right-hand side of Eq.(\ref{fscprim}) we can try to 
recover a form of scaling. Applying this procedure the
simplicity of the original Feynman idea is lost and the next correction
for the rise of the average transverse momentum could be introduced here as well. 
We have used in the present work 
the average transverse momentum rise of the form $\langle p_\bot \rangle\:=\:0.413\:-\:0.017\:\ln (s)\:+\:.00143\:\ln^2(s)$ 
shown in Fig.~4 of Ref.~\cite{cms7}.
The additional inelasticity control
parameter is an index in a power law multiplicative factor. 
These two modifications lead according 
to Eq.(\ref{fsc}) to only slightly more complicated scaling formula
\begin{equation}
{1 \over \sigma }~
{{d \sigma} \over {d \eta} }~
= \: \left( s \over s_0 \right)^{\alpha_F}
F_F \left( {2 \langle p_\bot \rangle \over \sqrt{s} }\: \sinh (\eta) \:\right)~~.
\label{fscprimptinel}
\end{equation}

\begin{figure}
\centerline{
\includegraphics[width=7.2cm]{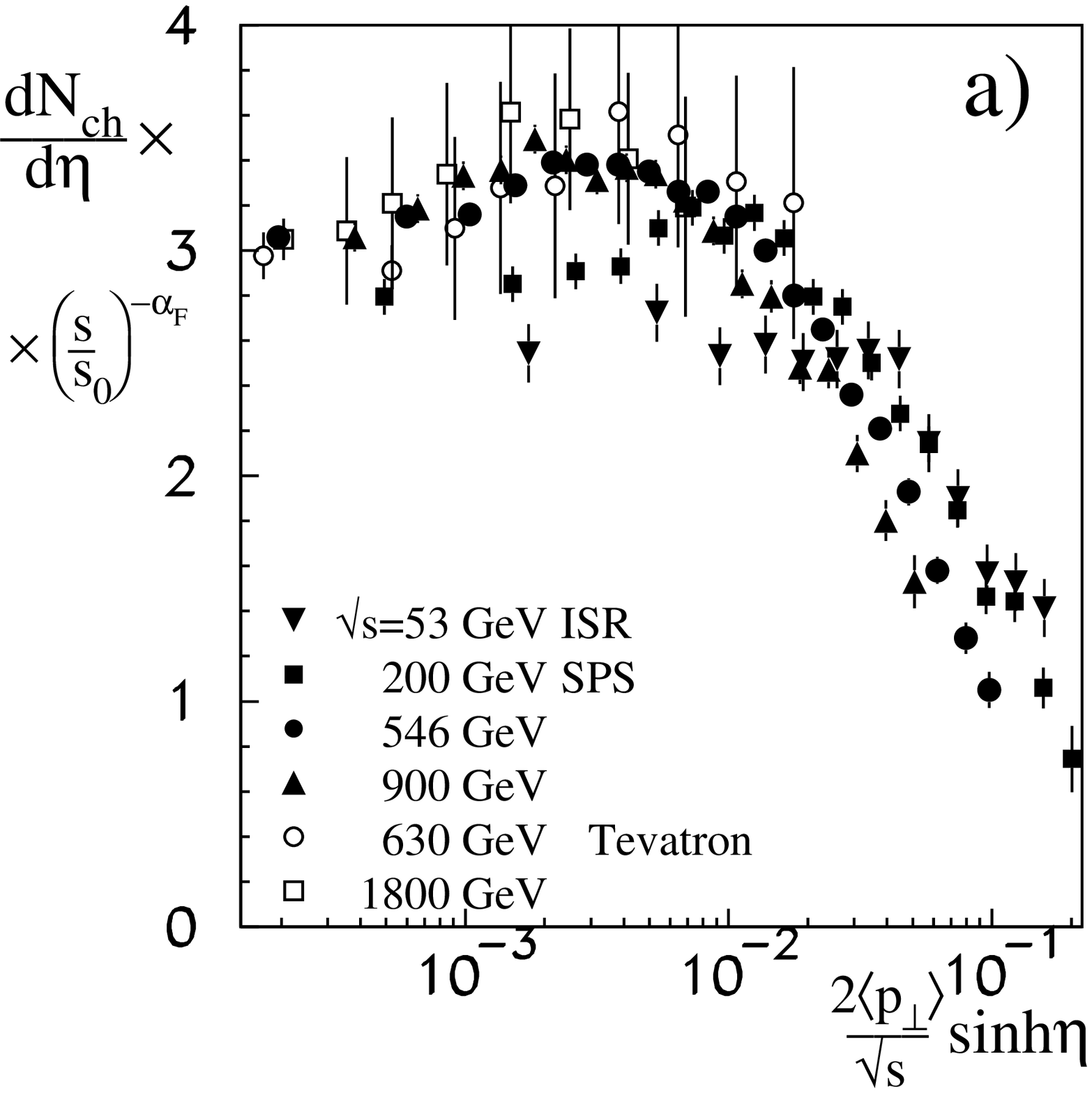}
\includegraphics[width=7.2cm]{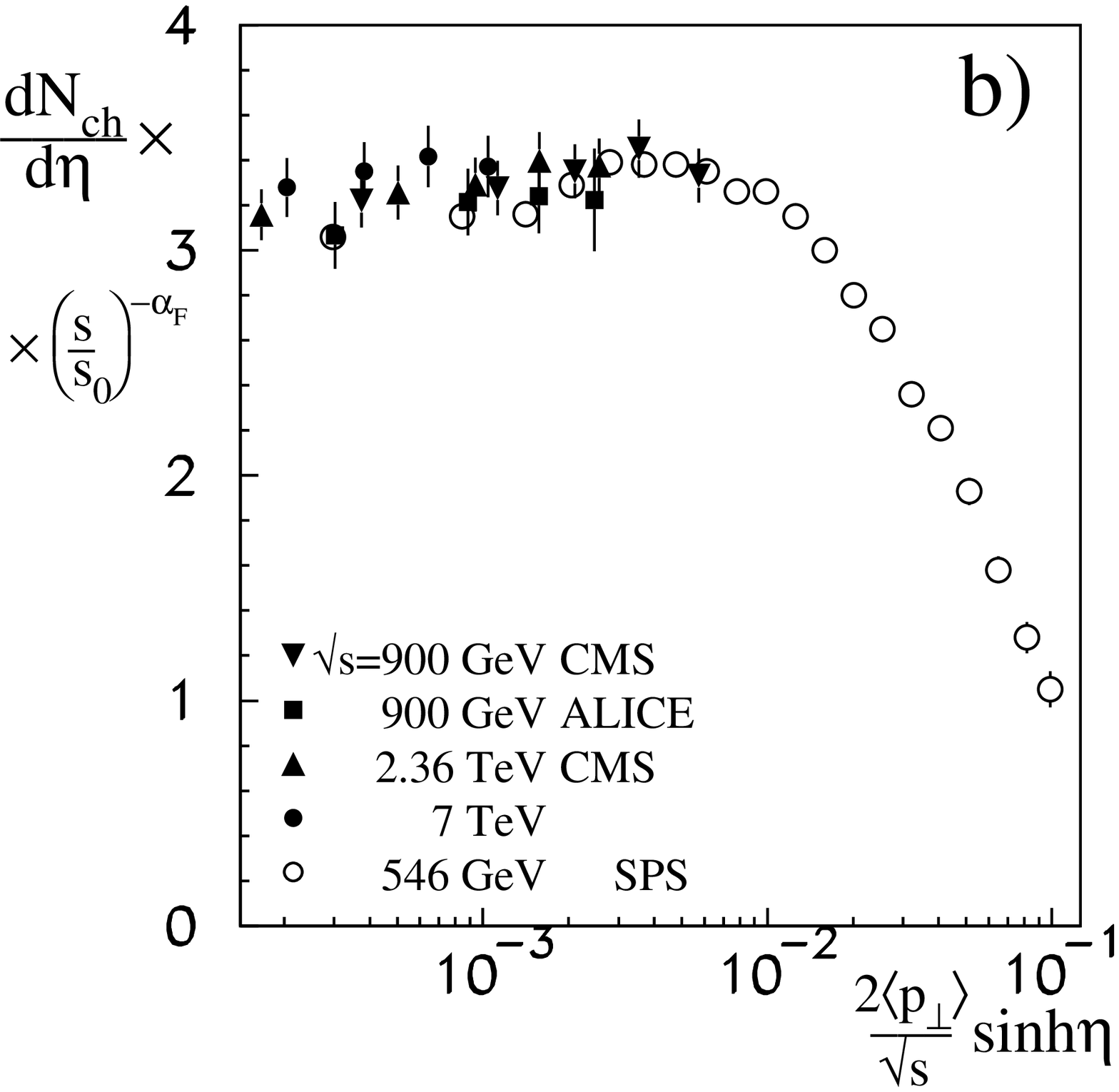}}
\caption{Pseudorapidity distributions 
shifted and transformed respectively adjusting $\alpha_F$ for ISR, SPS and Tevatron measurements (a), and distributions
measured by LHC experiments at energies from 900 GeV to 7 TeV compared with SPS 
$\sqrt{s}=546$~GeV UA5 result (b).}
\label{f2}
\end{figure}

We have used the UA5 data measured at $\sqrt{s_0} = 546$ GeV c.m.s. energy \cite{alner} 
as a datum.
The very accurate measured NSD 
pseudorapidity distribution have been used as a definition of the universal $F_F$ function. 
We adjusted the $\alpha_F$ parameter value to minimize the discrepancy between
Eq.(\ref{fscprimptinel}) scaling
prediction and the distributions of pseudorapidity measured at different energies: from ISR to 7 TeV of LHC.
The results are given in Fig.~\ref{f2}.

Values of $\alpha_F$ increase from $\sim 0.05$ found for ISR 53 GeV 
to  $\sim 0.11$ at LHC 7 TeV. 
The increase is statistically not very significant, 
at least for the overall inelasticity, what will be discussed later. 
The accuracy of the data scaling according to Eq.(\ref{fscprimptinel}) can be estimated with the help of
statistical tests. The $\chi^2$ values for the ISR and SPS are of about 
$\chi^2/NDF \approx 40/20$. The systematic uncertainties of the Tevatron and LHC results 
makes the $\chi^2/NDF$ smaller but the overall tendency seen in Fig.~\ref{f2} suggests 
strongly that proposed
modification of the Feynman scaling is not a right solution for 
the extrapolation of interaction properties to the very high interaction energies. 

\subsection{Wdowczyk and Wolfendale scaling}

It was shown in Ref.~\cite{twphlww} that the almost forty years old modification 
known as Wdowczyk and Wolfendale (WW) scaling \cite{ww} could 
be still satisfactory used to scale the interaction properties to the ultra high ($> 10^{19}$~eV) cosmic ray energies. 

The original idea of the WW scaling
\begin{equation}
f \left(x,\:p_\bot,\:s \right)~=~
(s/s_0)^\alpha \: f_{WW}\left( {x \: (s/s_0)^\alpha,\:p_\bot }\right)
\label{wwsc}
\end{equation}
is an extension of the Feynman fragmentation formula of 
Eq.~(\ref{xf}) (the limit for $\alpha=0$) with the possibility to get 
the 'thermodynamical limit' of $n \sim s^{1/4}$ with $\alpha = 0.25$.

The WW model in its version of mid '80 has been successfully used 
for the EAS studies around 'the knee'. Its extension introducing 
partial inelasticities (energy fraction carried by specific types of particles),
and the transverse momentum rise with interaction energy dependencies, as 
discussed above, gave better 
description of the production 
of different kinds of secondaries. As a result of this improvements 
the first power-law factor index was released and gave an extra model
parameter.
This more flexible 
formula was applied, e.g., in Ref.~\cite{alner} where
the agreement of the WW model predictions and the UA5 measured
rapidity distributions was shown. It should be mentioned that original 
Wdowczyk
and Wolfendale model gave a complete description of the multiparticle 
production process to be used mainly in EAS studies, so it
contains such details as partial inelasticities, transverse momenta, semiinclusive 
properties etc.
The fit shown in Ref.~\cite{alner} is the effective, average description of
inclusive data of rapidity (pseudorapidity) only. 

In the present work we explore the WW scaling of the form
\begin{equation}
{1 \over \sigma }~{{d \sigma} \over {d \eta} }~=~
{\left( s \over s_0 \right) ^{\alpha '} }~
F_{WW} \left( {\langle p_\bot \rangle \over \langle p_\bot^0 \rangle }\: \sinh (\eta)\:
{\left( s \over s_0 \right) ^{\alpha - 1/2} }\:\right)~~,
\label{wwfin}
\end{equation}
\noindent
where $\langle p_\bot^0 \rangle$
is the average transverse momentum at the datum interaction energy ($\sqrt{s_0}=546$ GeV).

\begin{figure}
\centerline{
\includegraphics[width=7.2cm]{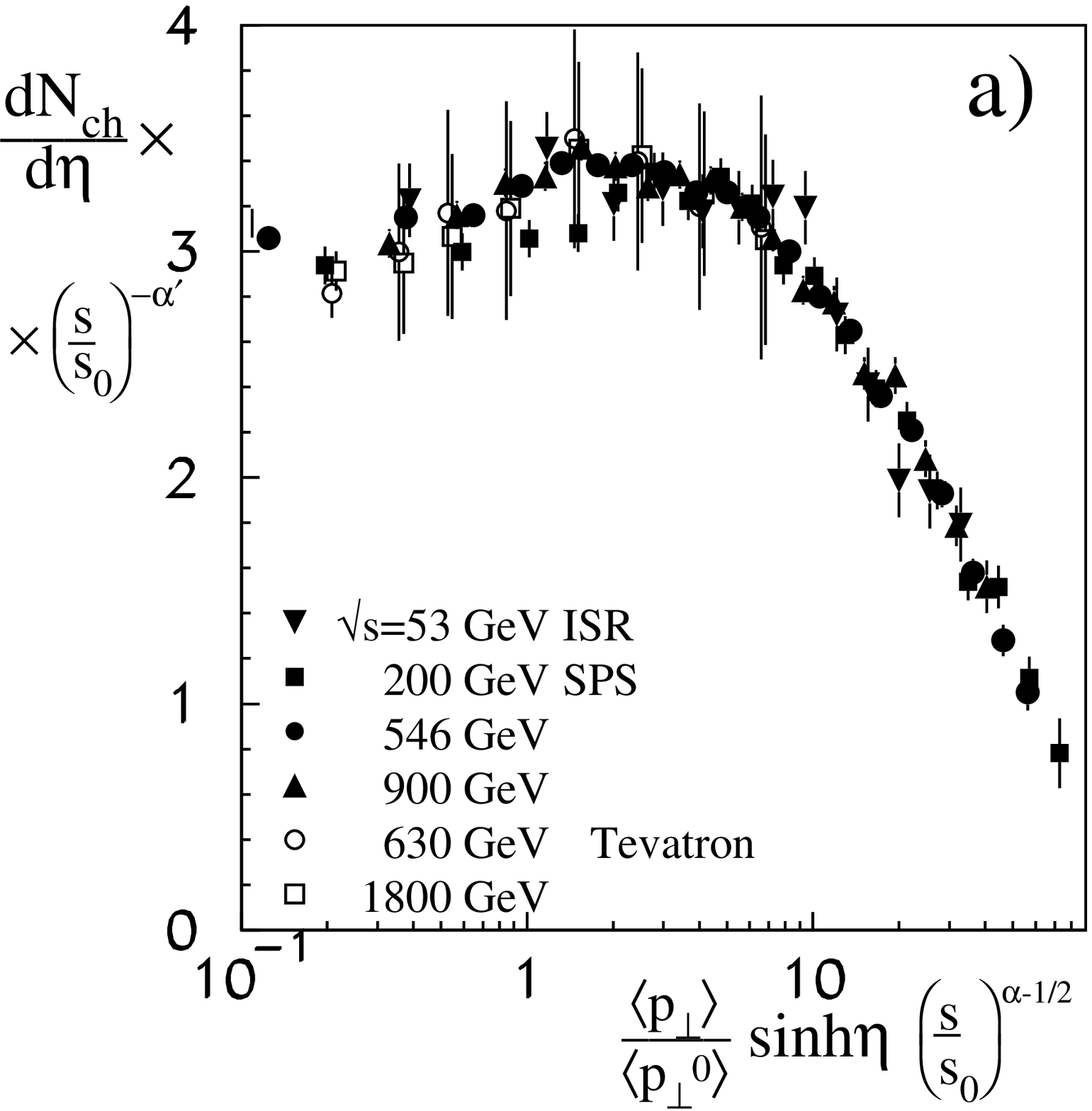}
\includegraphics[width=7.2cm]{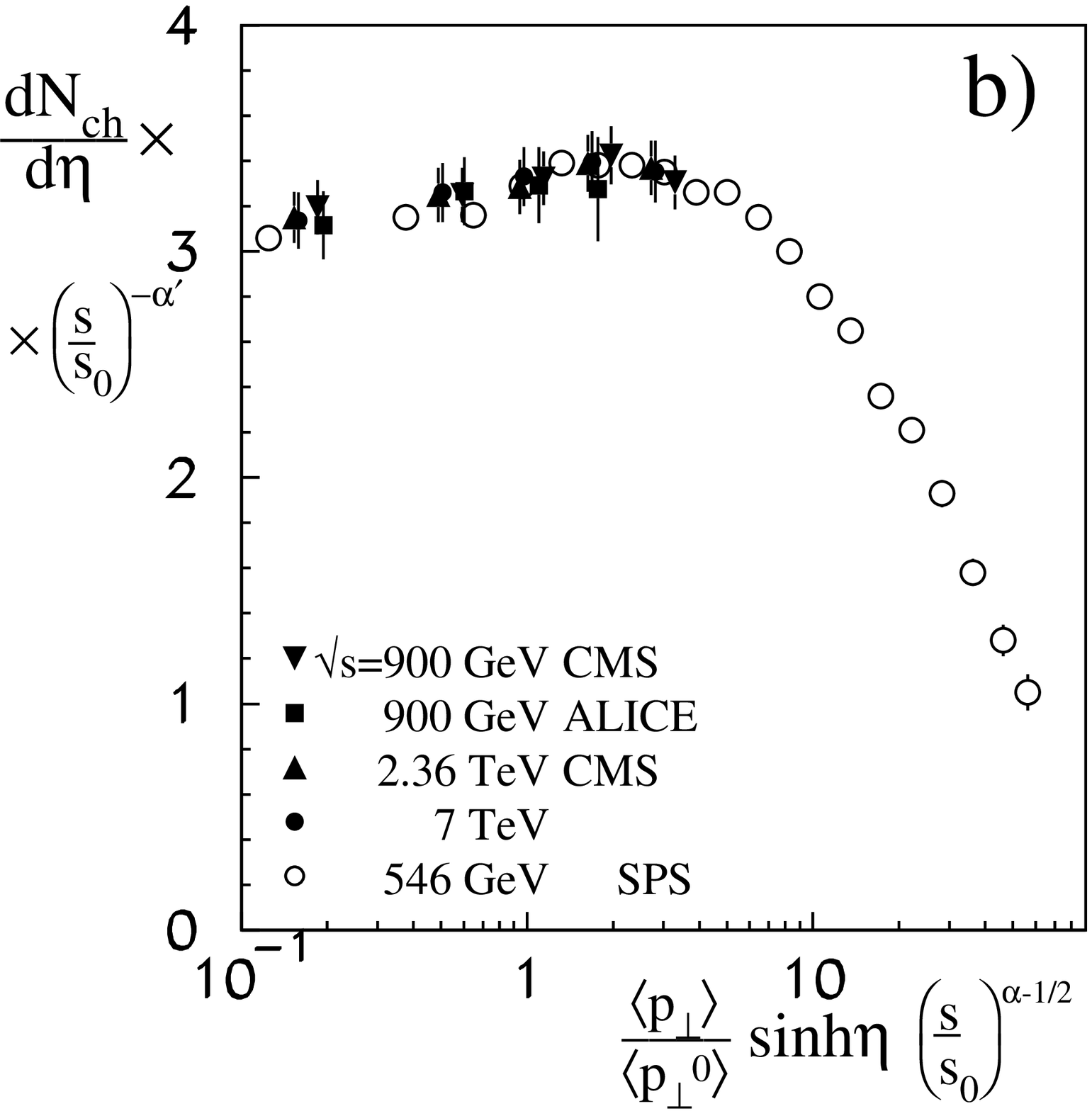}}
\caption{Wdowczyk and Wolfendale scaling 
with both parameters $\alpha$ and $\alpha '$ adjusted
 to each experimental data set.
\label{f3}}
\end{figure}

We have adjusted first both $\alpha$ and $\alpha '$ parameters independently to get the best scaling performance. Results are given in Fig.~\ref{f3}.

\begin{figure}
\centerline{
\includegraphics[width=7.2cm]{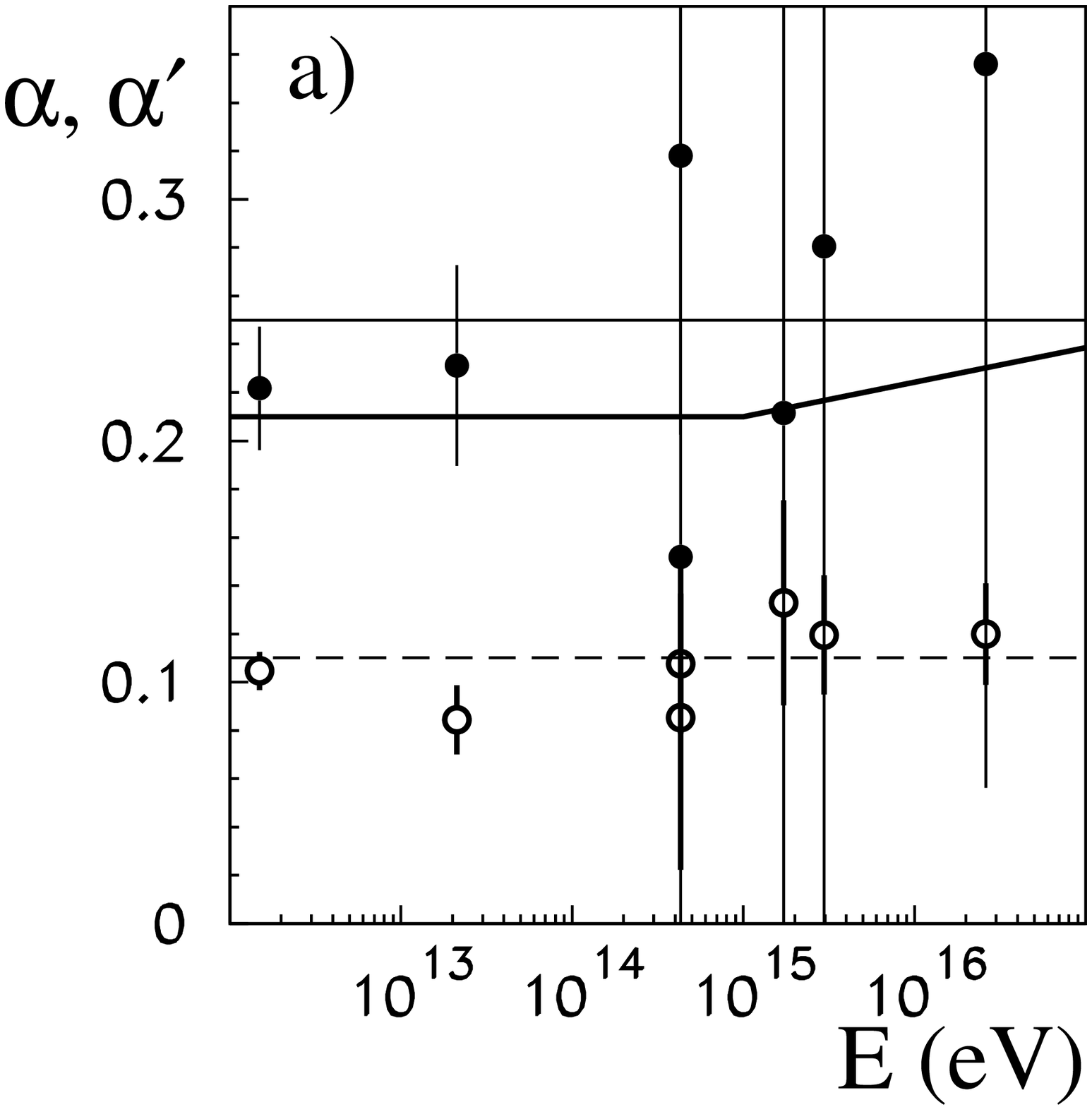}
\includegraphics[width=7.2cm]{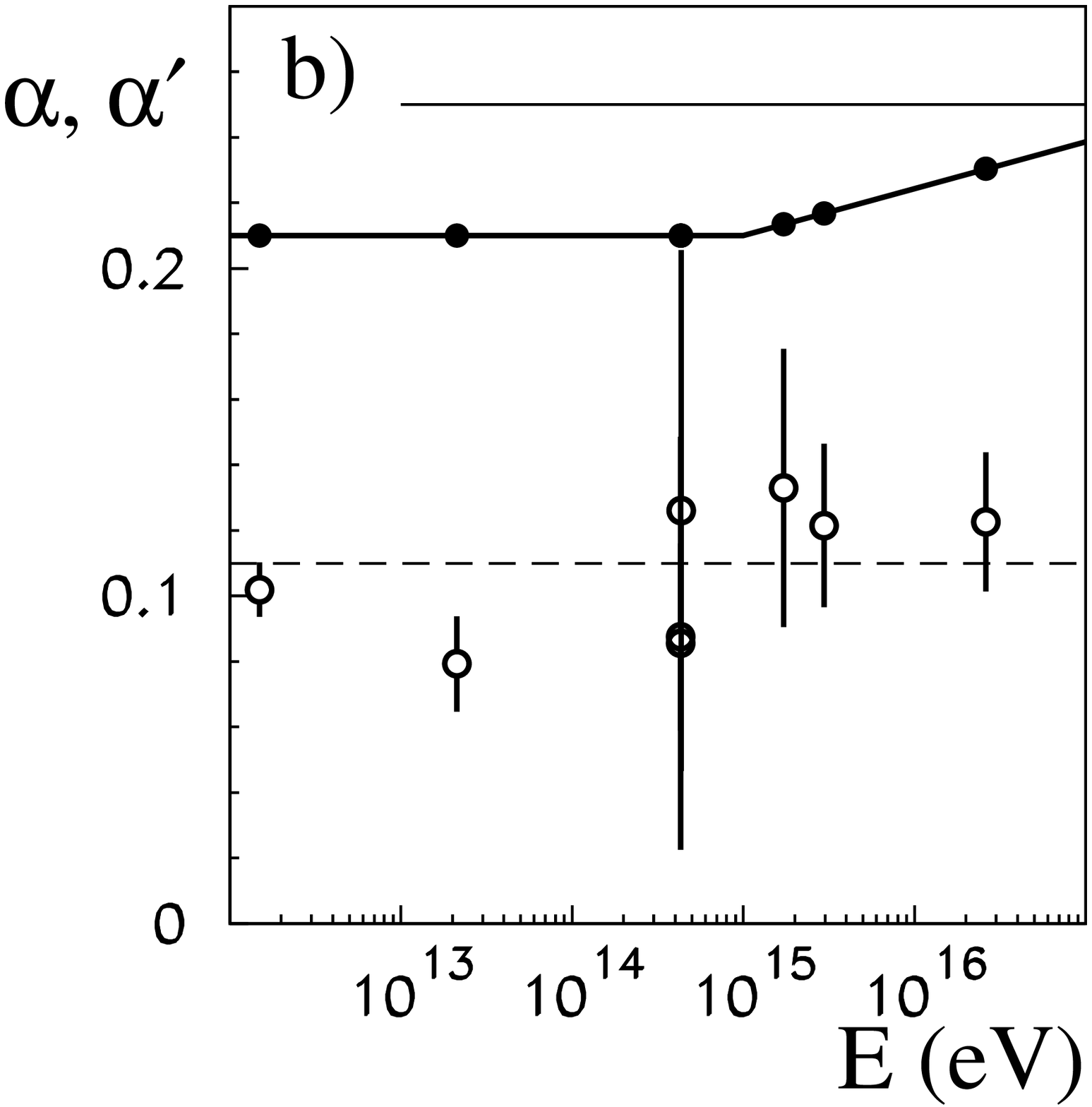}}
\caption{W{\&}W scaling parameters predictions for $\alpha$ (solid symbols and solid lines) and for $\alpha '$ 
(open symbols and dashed line) adjusted to the data (a), and values of
 $\alpha$ taken from the UHECR analysis \cite{twphlww} and only $\alpha '$ 
 used as a free parameter of the fit (b).
\label{f4}}
\end{figure}

Obtained values of $\alpha$ and $\alpha '$ are shown in Fig.~\ref{f4}a. Horizontal lines show 
results from Ref.~\cite{alner} (solid for $\alpha$ and dashed for $\alpha '$, 
respectively). The thick solid broken line is the result for $\alpha$ 
of our UHECR analysis \cite{twphlww}. 
It is seen that the predictions from Ref.~\cite{twphlww} and the LHC data 
are consistent. Although the
large uncertainties, which are result of limited rapidity range as well as 
possible systematics, do not
allow for any stronger conclusions. 

\begin{figure}
\centerline{
\includegraphics[width=7.2cm]{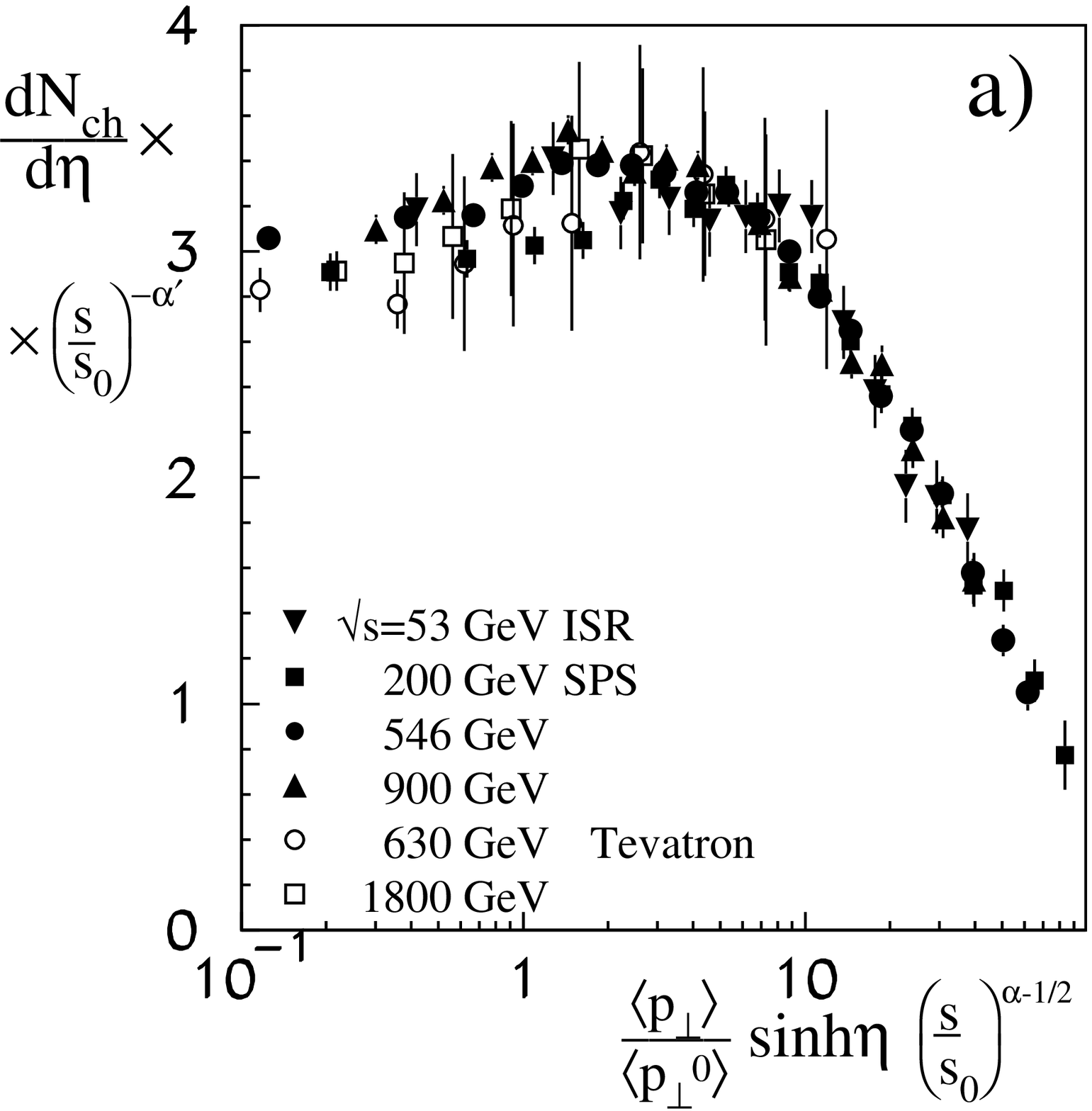}
\includegraphics[width=7.2cm]{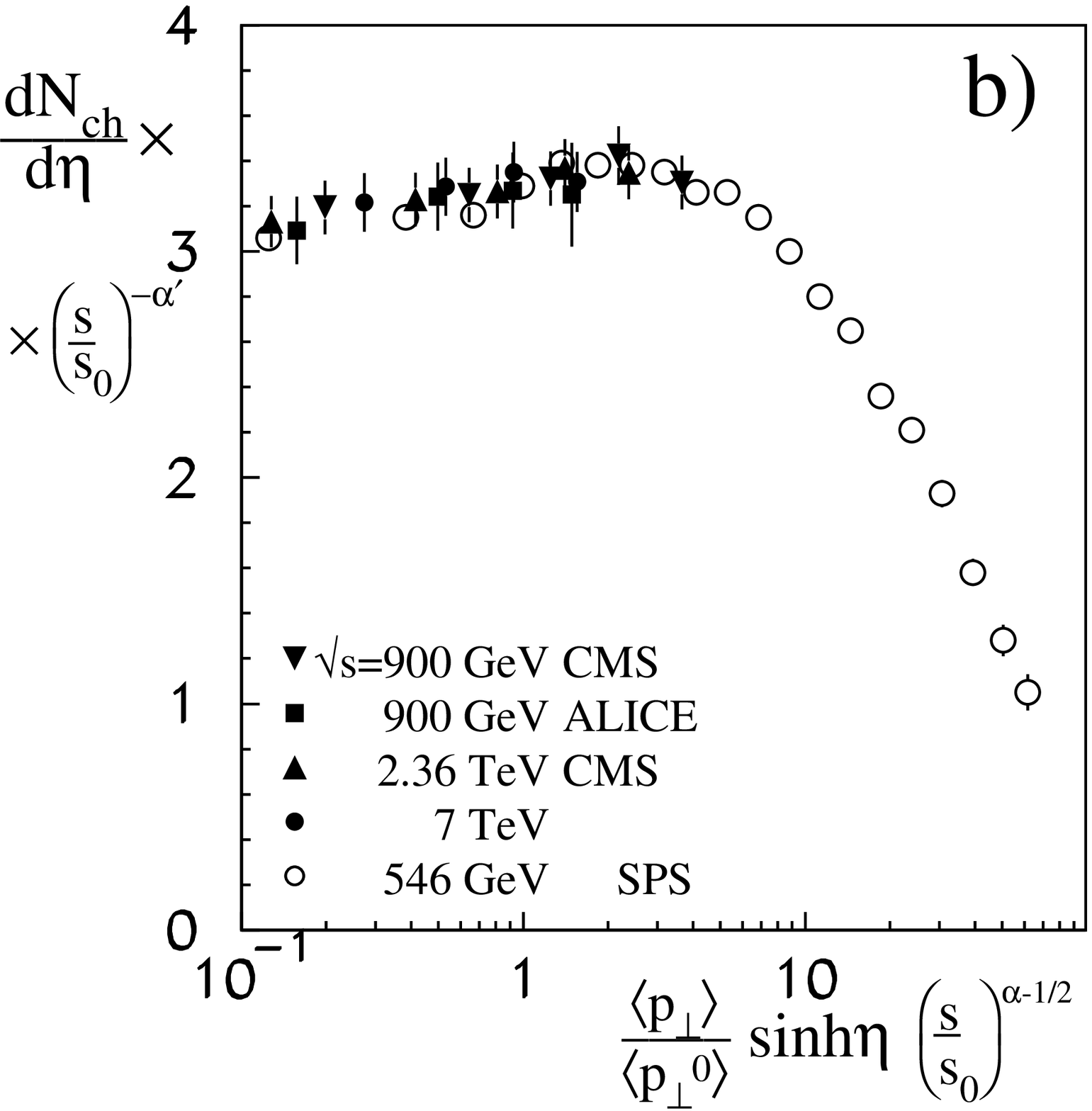}}
\caption{Wdowczyk and Wolfendale scaling results
with $\alpha$ set to the UHECR analysis data and $\alpha '$ adjusted
to each experimental data set shown as in the Fig.~\ref{f3}.}
\label{f5}
\end{figure}

We can, however, use the UHECR data analysis predictions 
for the values of $\alpha$ and test if results of the fit, with such reduced free 
parameter space, remains in agreement with the WW scaling.
It can be seen in Fig.~\ref{f5}

The data description is not much worst than the one presented in Fig.~\ref{f3}. 
The constancy of the $\alpha '$ suggested by WW original papers 
and seen in Fig.~\ref{f4}a, still holds as presented as 
in Fig.~\ref{f4}b.

\section{Inelasticity}

In Ref.~\cite{twphlww} it is found quite unexpected high energy behaviour of interaction
inelasticity coefficient. It was obtained as a result 
of the experimental suggestion that the composition of the UHECR 
is quite light, contains a significant proton fraction. 
The WW model with the strong Feynman scaling violation 
leads to 
continuous decrease of the energy fraction released to the secondaries produced
in very high energy interactions. 
Eq.(\ref{wwfin}) gives the inelasticity energy dependence
\begin{equation}
K(s)~=~K_0\:\left(s \over s_0 \right) ^{(\alpha ' - \alpha)}~,
\label{kinelww}
\end{equation}
\noindent
while for the modified Feynman scaling formula Eq.(\ref{fscprimptinel}) it is
\begin{equation}
K(s)~=~K_0\:\left(s \over s_0 \right) ^{\alpha_F}~.
\label{kinelf}
\end{equation}

\begin{figure}
\centerline{
\includegraphics[width=9cm]{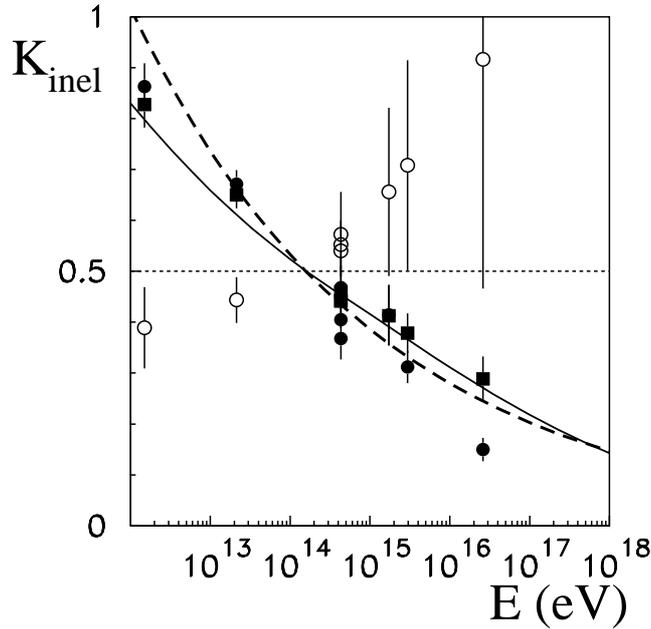}}
\caption{Inelasticity calculated with WW scaling assumption (filled symbols - 
circles for both $\alpha$ and $\alpha '$ adjusted (Fig.\ref{f3}) and squares 
for UHECR inspired $\alpha$ (Fig.~\ref{f5}).
\label{f6}}
\end{figure}

\noindent
In the Fig.~\ref{f6} we have shown results of our analysis. Open symbols show the fast rise 
of the inelasticity for modified Feynman scaling formula. Even if the $\alpha_F$ follow 
the lower energy, smaller value, in the UHECR domain the saturation is expected.  
Filled symbols were obtained for WW scaling.
The solid line gives the
predictions from Ref.~\cite{twphlww} obtained using UHECR data. The dashed line is the 
fit from Ref.~\cite{alner} of the WW scaling parameters to SPS data. 
The value of 0.5 is also shown.

The open symbols are for the modified Feynman scaling with $\alpha_F$ parameter.
Solid line shows the UHECR data analysis prediction from Ref.~\cite{twphlww}.
Dashed line is the inelasticity fit from Ref.~\cite{alner}. The 'canonical' value 0.5 is
shown by short dashed line.

\section{Summary}

We have shown that the minimum bias pseudorapidity distributions measured by LHC
experiments can be very well described with the scale-breaking Wdowczyk and Wolfendale 
formula. 

The scaling violation observed for the energies up to SPS $\sqrt{s}=900$~GeV 
and 1800 GeV in Tevatron was
uphold recently in the analysis of new UHECR data.

The phenomenological model of Wdowczyk and Wolfendale introduces two 
model parameters. The value of one of them: $\alpha$, was originally found to be equal to 
0.13 using interpolation 
of the \mbox{$x_F=p_\| /p_{\rm max}$} distributions between 
$\sqrt{s} \approx 10$ GeV and ISR energies. Later interpolations including SPS
data gave the value of 0.18 and finally the effective value of 0.25 was found in Ref.~\cite{alner}.
The increase of the central rapidity density reported also in Ref.~\cite{alner} suggests $\alpha 
= 2 \times 0.105 = 0.21$. This value gives the Extensive Air Showers development 
maximum position $x_{\rm max}$ for proton initiated showers not far 
from measured \cite{pao,hires} as it is shown in Ref.~\cite{twphlww}.

The UHECR data
suggests further smooth
rise of the scale-breaking parameter. 
The first measurements at LHC up to 7 TeV c.m.s. energy 
agree with the trend observed at lower energies and seems to
smoothly bridge accelerator results and these on very high energy 
interaction of cosmic ray protons.
The limited range of measured pseudorapidities does not allow for a stronger statement. 
The more forward particle production data is highly welcome.

The rising inelasticity for (modified) Feynman scaling is obviously in contrary to the 
Wdowczyk and Wolfendale scaling and cosmic ray data. 
Comparing the pseudorapidity distributions in Figs.~\ref{f3}b and
\ref{f5}b we can say that the LHC pseudorapidity data analysis favours
the second possibility.





\bibliographystyle{elsarticle-num}
\bibliography{<your-bib-database>}

\begin{thebibliography}{00}

\bibitem{gzk}K.~Greisen, Phys. Rev. Lett. {\bf 16}, 748 (1966);\\
G.~T. Zatsepin and V.~A. Kuz'min, J. Exp. Theor. Phys. Lett.{\bf 4}, 78 (1966).
\bibitem{pao}F. Sch\"{u}ssler for the Pierre Auger Collaboration,
Proc. 31st ICRC {\L }\'{o}d\'{z}, Poland, 2010.
\bibitem{hires}P. Sokolsky and G.B. Thomson, J. Phys. G 
{\bf 34}, R401 (2007).
\bibitem{limi-fra}J. Benecke, T. T. Chou, C. N. Yang, and E. Yen, Phys. Rev. 
{\bf 188}, 2159 (1969).
\bibitem{feynman}R.P. Feynman, Phys. Rev. Lett. {\bf 23}, 
1415 (1969).
\bibitem{ww}J. Wdowczyk and A. W. Wolfendale, Nature 
{\bf 236}, 29 (1972); \\
J. Wdowczyk and A.W. Wolfendale, J. Phys. A {\bf 6} 
L48 (1973);\\
J. Olejniczak, J. Wdowczyk and A.W. Wolfendale,  J. 
Phys. G. {\bf 3} 847(1977);\\
J. Wdowczyk and A.W. Wolfendale, Nuovo Cim {\bf 54}A 
433 (1977);\\
J. Wdowczyk and A.W. Wolfendale, Nature {\bf 306}, 
347 (1983); \\
J. Wdowczyk and A.W. Wolfendale, J. Phys. G {\bf 10}, 
257 (1984).
\bibitem{twphlww}T. Wibig, Phys. Lett. B {\bf 678} 60, (2009).
\bibitem{alice} K. Aamodt et al., [ALICE Collaboration] 
Eur. Phys. J. C{\bf 65}, 111
(2010).
\bibitem{cms900}
CMS Collaboration. JHEP {\bf 2}, 41 (2010)

\bibitem{cms7}V. Khachatryan et al., (CMS Collaboration),
Phys. Rev. Lett. {\bf 105}, 022002 (2010).
\bibitem{atlas}
ATLAS Collaboration,  Phys. Lett. B{\bf 688}, 21
(2010).
\bibitem{alner}G.J. Alner {\it et al.} (UA5 Collaboration),
Zeitschrift fur Physik C {\bf 33}, 1 (1986).
\bibitem{twjpgpt}T. Wibig, J. Phys. G {\bf 37}, 115009 (2010).



\end{thebibliography}



\end{document}